\documentclass[aps,prc,preprint,showpacs]{revtex4}
\usepackage{hyperref}
\usepackage{graphicx}
\usepackage{bm}

\newcommand{\bra}{\langle}
\newcommand{\ket}{\rangle}

\begin{document}

\title{Model Dependence of the  $^2$H Electric Dipole Moment}

\author{Iraj R. Afnan}
\affiliation{School of Chemical and Physical Sciences\\
      Flinders University, GPO Box 2100, Adelaide 5001, Australia     }
      \email{Iraj.Afnan@Flinders.edu.au}
\author{Benjamin F. Gibson}
\affiliation{Theoretical Division, Los Alamos National Laboratory \\
Los Alamos, NM 87545, USA}
\email{bfgibson@lanl.gov}

\date{\today}

\begin{abstract}
\textbf{Background:} Direct measurement of the electric dipole moment (EDM) of the neutron 
lies in the future; measurement of a nuclear EDM may well come first.  The deuteron is one 
nucleus for which exact model calculations are feasible. 
\textbf{Purpose:} We explore the model dependence of deuteron EDM calculations. 
\textbf{Methods:} Using a separable potential formulation of the Hamiltonian, we examine the
sensitivity of the deuteron EDM to variation in the nucleon-nucleon interaction.  We write
the EDM as the sum of two terms, the first depending on the target wave function with plane-wave intermediate states, and the second depending on intermediate multiple scattering in the $^3$P$_1$ channel, the latter 
being sensitive to the off-shell behavior of the $^3$P$_1$ amplitude.
\textbf{Results:} We compare the full calculation with the plane-wave approximation result, 
 examine the tensor force contribution to the model results, and explore the effect of short 
range repulsion found in realistic, contemporary potential models of the deuteron.
\textbf{Conclusions:} Because one-pion  exchange dominates the EDM calculation, separable 
potential model calculations will provide an adequate description of the $^2$H EDM until
such time as a better than 10\% measurement is obtained.
\end{abstract}

\maketitle

\section{Introduction}
With the discovery of parity ($P$) violation, which was suggested by Lee and Yang~\cite{LY56}, 
Landau~\cite{La57} deduced that charge conjugation and parity ($CP$) invariance implies that 
the electric dipole moment (EDM) of particles, \textit{e.g.} the neutron, should be zero.  If the 
$CPT$ theorem is valid, which is the case for gauge theories, then any $CP$ violation would
also imply a corresponding time reversal ($T$) invariance violation.   Predating the discovery
of parity violation in the weak interaction, Purcell and Ramsey~\cite{PR50} had pointed out that
there was lacking any experimental test of parity conservation in the strong interaction.  With
their student Smith~\cite{SPR57} they set limits on the EDM of the neutron of the order of 
$d_n < 5 \times 10^{-20}\, e \, cm$.   The Standard Model of fundamental interactions 
predicts values for EDMs (due to second order W boson exchange) which are significantly smaller 
than contemporary experiments can  detect, of the order of $10^{-31}\, e \, cm$.  Therefore, an 
unambiguous observation of a nonzero EDM at current capabilities would imply a yet to be 
discovered source of $CP$ violation~\cite{He95,KL97}.  The new physics could arise in the strong 
interaction sector (\textit{e.g.}, the $\theta$ term), or in the weak interaction sector [\textit{e.g.}, 
Super Symmetric models or Left/Right (boson mass) symmetry breaking].   Current limits on the 
nucleon EDM are of the order of $10^{-26}\, e \, cm$.  Even were one to establish a nonzero 
neutron and proton EDM, those two results would at best determine the isoscalar and isotensor 
components but would not isolate any isovector component.  Thus, one would need a third 
measurement, such as the deuteron EDM, to fully elucidate the isospin nature of the EDM operator.  
Both $PT$ violating and $P$ conserving, $T$ violating potentials may give rise to an EDM~\cite{He95}, 
but one-pion exchange contributes only to the former.  We concentrate here upon the effects 
due to $PT$ invariance violation in the nuclear potential.

The deuteron is attractive as the focus of an EDM investigation, both theoretically 
and experimentally, because a method has been proposed to directly measure the EDM of 
charged ions in a storage ring~\cite{Kh98, Fa04, Se04, Or06}.  A permanent EDM can arise 
because a $PT$ violating interaction can induce a small P-state admixture in the deuteron wave 
function, one which produces a non vanishing matrix element of the charge dipole operator 
$\tau^z_{-} e\vec{r}$.  Although this two-body EDM contribution must be disentangled from 
the one-body contributions of the neutron and proton, the neutron and proton EDMs tend to 
cancel in the case of the isospin zero $^2$H.  (If the nucleon EDM were a pure isoscalar as is 
the case in the $\theta$ model, then this cancellation would be exact.)  Therefore, the $PT$ 
violating nucleon-nucleon (\textit{NN}) interaction can contribute significantly to the deuteron 
EDM.  Because the deuteron is reasonably understood and has been accurately modeled, reliable 
calculations are possible.  Our purpose is to address the sensitivity of the deuteron EDM to 
the nuclear physics in the modeling of the nucleon-nucleon interaction.  Beyond 
understanding the model dependence of the $^2$H EDM, our goal is to determine an 
appropriate model approximation with which one might reliably calculate the nuclear physics 
contribution to the $^3$He and $^3$H EDMs.  Therefore, we examine the uncertainties in the 
deuteron EDM calculation arising from the short range repulsion in the ground state wave 
function, the dependence on the size of the deuteron D-state, and the properties of the 
$^3$P$_1$ continuum in intermediate states.  

For the purpose of completeness and to place our work in context, we note that 
Avishai~\cite{Av85} first estimated the two-body deuteron EDM [ see Eq.~(\ref{eq:2}) ] $d_D^{(2)}$
using a separable potential model due to 
Mongan~\cite{Mo69}.  He reported a value of $- 0.91\ A\,e \,fm$ when he utilized 
the physical pion mass for the exchanged meson.  [Note: To exclude the $PT$ violating and strong coupling constants in the one pion exchange nucleon-nucleon interaction for the quoted values of the EDM, we have introduced $A=\bar{g}^{(1)}_{\pi NN}\,g_{\pi NN}/(16\pi)$.]
  However, there is an ambiguity in Avishai's results, in that he states 
his final result in terms of $A/2$.  Because the particular separable potentials used 
by Avishai were not specified, we were unable to fully confirm his reported numbers.   
Khriplovich and Korkin~\cite{KK00} later estimated $d_D^{(2)}$ using a zero-range 
approximation in the chiral limit ($m_\pi \rightarrow 0$) and obtained a value of $- 0.92\  
A\, e\, fm$.  This result does not depend upon the $^3$P$_1$ interaction and should, 
therefore, be directly comparable to our `plane wave' result.  Finally, using the Argonne and 
Nijmegen contemporary realistic potential models A$v_{18}$, Reid93, and Nijm II~\cite{SKTS94} 
Liu and Timmermans~\cite{LT04} obtained for the polarization component of the 
two-body contribution to the deuteron EDM $d_D^{(2)}$ values of $- 0.72\ A\, e\, fm$, 
$- 0.73\ A\, e\, fm$, and $- 0.74\ A\, e\, fm$,  respectively.  These relatively 
model-independent results suggest that pion exchange is indeed the essential aspect 
of the model.  The differing degree of softness of the three potentials at intermediate 
range correlates with the values for $d_D^{(2)}$, the Nijm II potential being the softest 
and producing the largest EDM.  The important conclusion for our purpose is that all 
three models yield essentially the same result; within the range of uncertainty defined 
by the three models utilized, the value of the polarization component of $d_D^{(2)}$ 
can be said to be $\approx - 0.73\pm .01\ A\, e\, fm$.  Moreover, Liu and 
Timmermans estimated that the meson exchange current contribution was substantially 
smaller, calculated to be less than 5\% of the potential model contribution.  In any case, 
our goal is to determine an appropriately simple model with which one can calculate 
reliably the $^2$H, $^3$He, and $^3$H EDMs, so that our numerical comparisons will 
be made with the $- 0.73\pm .01\ A\, e\, fm$ value.

\section{Nucleon contributions}
The total one-body contribution $d_D^{(1)}$ to the deuteron EDM due to the neutron and 
proton is the sum of the individual nucleon EDMs:
\begin{equation}
                d_D^{(1)} = d_n + d_p \; ,  \label{eq:1}
\end{equation}
whereas the total deuteron EDM is the sum of this one-body contribution and the two-body
contribution $d_D^{(2)}$,
\begin{equation}
                d_D = d_D^{(1)} + d_D^{(2)} = (d_n + d_p) + d_D^{(2)} \; .\label{eq:2}
\end{equation}
As has been noted, the neutron and proton EDMs can arise from a variety of sources.   
Because we have nothing new to add to prior analyses of the nucleon EDM, we 
adopt the approach advanced by Liu and Timmermans~\cite{LT04}:
\begin{equation}
  d_D^{(1)} \simeq  0.22\times10^{-2} \bar{G}_\pi^{(1)} 
                       + O(\bar{G}_\pi^{(0,2)},\bar{G}_{\rho,\omega,\eta}) \; ,\label{eq:3}
\end{equation}
which is expressed in terms of $\bar{G}_X^{(i)}$, the product of the strong coupling 
constant $g_{XNN}$ and the associated $PT$ violating meson-nucleon coupling constant 
$\bar{g}_X^{(i)}$.  (For example, $\bar{G}_\pi^{(1)} = \bar{g}_{\pi NN}^{(1)} \, g_{\pi NN}$.)   As 
noted in Ref.~\cite{LT04}, the contributions from the neutron and proton EDMs have a sizable 
theoretical uncertainty, but the significant cancellation between $d_n$ and $d_p$ is clear.  
For the two-body contribution to $d_D^{(2)}$ the mean value obtained by Liu and 
Timmermans can be expressed as
\begin{equation}
      d_D^{(pol)} = 1.45\times10^{-2}\bar{G}_\pi^{(1)}  \; ;\label{eq:4}
\end{equation}
this corresponds to the EDM value of $- 0.73\ A\, e\, fm$.   Hence, for the deuteron there can be
little doubt that the nuclear physics contribution to $d_D^{(2)}$ dominates.  Even an uncertainty 
of 50\% in $d_D^{(1)}$ contributes only in a minor way.  It is the nuclear model aspects of 
the $d_D^{(2)}$ dominant term in the $^2$H EDM that we investigate below in detail.

\section{Two-body Contributions}

The interaction Hamiltonian for the ground state of the system consists of two components: 
(i)~The strong interaction component $v$ based on nucleon-nucleon potentials with 
parameters adjusted to fit the experimental phase shifts. (ii)~The $PT$ violating component 
$V$ which we parametrize in terms of one pion exchange (OPE) with one strong interaction 
vertex $g_{\pi NN}$ and a $PT$ violating vertex $\bar{g}^{(1)}_{\pi NN}$.  As a result our 
Hamiltonian takes the form
\begin{equation}
H= H^S + H^{PT}\quad\mbox{where}\quad H^S= H_0 + v \quad\mbox{and}\quad H^{PT}=V\ .\label{eq:5}
\end{equation}
Because  $H^{PT}$ will mix different parity states, i.e., for the deuteron we get coupling
between the $^3$S$_1$-$^3$D$_1$ large component and the $^3$P$_1$ small component, 
we can write the Schr\"odinger equation for the Hamiltonian in Eq.~(\ref{eq:5})  
\begin{equation}
H\,|\Psi\ket = E\,|\Psi\ket \label{eq:6}
\end{equation}
as a set of coupled equations of the form
\begin{eqnarray}
(E-H_0)\,|\Psi_L\ket &=& v\,|\Psi_L\ket + V\,|\Psi_S\ket\ \label{eq:7}\\
(E-H_0)\,|\Psi_S\ket &=& v\,|\Psi_S\ket + V\,|\Psi_L\ket \ ,\label{eq:8}
\end{eqnarray}
where the total wave function is the sum of the large and small components:
$|\Psi\ket = |\Psi_L\ket + |\Psi_S\ket$. 

Because $V\ll v$, we have that $V\,|\Psi_S\ket\ll v\,|\Psi_L\ket$, and we can, to a good 
approximation, write Eq.~(\ref{eq:7}) as
\begin{equation}
(E-H_0)\,|\Psi_L\ket = v\,|\Psi_L\ket\ ,\label{eq:9}
\end{equation}
which is the Scr\"odinger equation for the ground state of the system in the absence of the 
$PT$ violating interaction.  On the other hand, the small component of the wave function $|\Psi_S\ket$
is given by the solution of Eq.~(\ref{eq:8}) in terms of the amplitude $t(E)$ for the strong potential 
$v$ as 
\begin{equation}
|\Psi_S\ket = G(E)\,V\,|\Psi_L\ket\quad\mbox{with}\quad 
G(E) = G_0(E) + G_0(E)\,t(E)\,G_0(E)\ ,\label{eq:10}
\end{equation}
where $G_0(E)=(E-H_0)^{-1}$ is the free Green's function, and $t(E)$ is the amplitude 
in the partial wave of the small component of the wave function, e.g., for the 
deuteron $t(E)$ is the amplitude in the $^3$P$_1$ partial wave at the ground state energy.

Because the dipole operator 
\begin{equation}
O_d = \frac{e}{2}\ \sum_{i} \vec{r}_i\ \tau_z(i) \label{eq:11}
\end{equation}
is odd under parity, we can write the two-body deuteron EDM ($d_D^{(2)}$) in terms of 
the total ground state wave function $|\Psi\ket = |\Psi_L\ket+|\Psi_S\ket$ as
\begin{equation}
d_D^{(2)} = \bra\Psi|\,O_d\,|\Psi\ket = \bra\Psi_L|\,O_d\,|\Psi_S\ket + 
                   \bra\Psi_S|\,O_d\,|\Psi_L\ket\ ,\label{eq:12}
\end{equation}
where the matrix element of the dipole operator between the small and large component 
of the wave function can be written in terms of the charge $e$ and the constant $A$ as
\begin{eqnarray}
\bra\Psi_L|\,O_d\,|\Psi_S\ket &=& \bra\Psi_L|\,O_d\,G_0(E)\,V\,|\Psi_L\ket 
+ \bra\Psi_L|\,O_d\,G_0(E)\,t(E)\,G_0(E)\,V\,|\Psi_L\ket\ \label{eq:13}\\
&\equiv& \frac{e}{2}\,\left[d_{PW} + d_{MS}\right]\,A \quad\mbox{with}\quad 
A\equiv\frac{\bar{g}^{(1)}_{\pi NN}\,g_{\pi NN}}{16\pi}\ .\label{eq:14}
\end{eqnarray}
In Eq.(\ref{eq:13}) the first term on the right hand side (rhs) involves a complete set 
of intermediate plane wave states and is, up to a constant, 
the `plane wave' contribution 
$d_{PW}$. The second term on 
the rhs of Eq.~(\ref{eq:13}) involves multiple scattering via the amplitude $t(E)$ and 
is the `multiple scattering' contribution $d_{MS}$.  One should note that $E<0$ is the ground state energy, and as 
a result we need the amplitude $t(E)$ at an unphysical point corresponding to the $^2$H
bound state energy.

\section{Numerical results}

The primary motivation for the present investigation is: (i)~to determine the sensitivity of 
$d_D^{(2)}$ to properties of the deuteron, \textit{e.g.} the $D$-state probability and the 
short range behavior of the deuteron wave function. (ii)~to determine the relative importance 
of $d_{PW}$ and $d_{MS}$. This will suggest the significance of multiple scattering terms as 
one proceeds to heavier nuclei. (iii)~The role of the $^3$P$_1$ interaction in determining the magnitude of $d_{MS}$ and therefore the appropriateness of the $d_{PW}$ approximation in heavier nuclei. Before we proceed to illustrate the sensitivity of the deuteron EDM to nuclear structure effects due to the nuclear interaction, we should detail our choice of nucleon-nucleon interactions and their fit to those aspects of the two-body data relevant to the determination of the EDM.

\subsection{Two-body potentials}

The input two-body interactions consists of: (i)~The $PT$ violating one pion exchange potential. (ii)~The deuteron wave function in the absence of the $PT$ violating interaction. (iii)~The $^3$P$_1$ interaction that couples to the deuteron $^3$S$_1$-$^3$D$_1$ potential as a result of the introduction of the $PT$ violating potential. The choice of these interactions is motivated by the questions raised regarding the sensitivity of the EDM to nuclear structure effects and the hope of extending the analysis to $^3$H and $^3$He using the $d_{PW}$ approximation.

For the $PT$ violating interaction we have chosen the standard isovector one-pion exchange given by~\cite{PH92}
\begin{equation}
V = -A\left[(\vec{\sigma}^{(-)}\cdot\hat{r})\,\tau_z^{(+)} + (\vec{\sigma}^{(+)}\cdot\tau_z^{(-)}\right]\,f(r)\ ,\label{eq:15}
\end{equation}
where the radial dependence is given by
\begin{equation}
f(r) = - \frac{1}{m_\pi}\,\frac{d}{dr}\left(\frac{e^{-m_\pi r}}{r}\right)\ ,\label{eq:16}
\end{equation}
with $m_\pi$ being the pion mass. Here we have combined the strength of the strong and $PT$ violating vertices in the constant $A$ given in Eq.~(\ref{eq:14}). This allows us to express the numerical value of the EDM in terms of $A\,e$ with $e$ the charge on the proton. Finally, the spin and isospin operators in Eq.~(\ref{eq:15}) are given by $\vec{\sigma}^{(\pm)} = (\vec{\sigma}^{(1)}\pm\vec{\sigma}^{(2)})$ and $\tau_z^{(\pm)} = (\tau_z^{(1)}\pm\tau_z^{(2)})$.

The strong $^3$S$_1$-$^3$D$_1$ interaction basically defines the deuteron wave function. Here we resort to a separable representation of the interaction to simplify the computation when we proceed to the EDM for the three-nucleon system. As a result the partial wave expansion of the strong interaction in momentum space is written as 
\begin{equation}
   \bra\vec{k}|\,v\,|\vec{k}'\ket = \sum_{Sjtm}\sum_{\ell\ell'}\ \bra\hat{k}|{\cal Y}^t_{(\ell S)jm}\ket\ 
   v^{Sjt}_{\ell\ell'}(k,k')\ \bra{\cal Y}^t_{(\ell'S)jm}|\hat{k}'\ket\ ,  \label{eq:17}
\end{equation}
with $|{\cal Y}^t_{(\ell S)jm}\ket$ eigenstates of the orbital angular momentum $\ell$, spin $S$, total angular momentum $j$ and isospin $t$. The separability of the potential is defined by the requirement that
\begin{equation}
v^\alpha_{\ell\ell'}(k,k') = g^\alpha_\ell(k)\ \lambda^\alpha_{\ell \ell'}\ g^\alpha_{\ell'}(k')\ ,\label{eq:18}
\end{equation}
where $\alpha = (Sjt)$.
Here we wish to examine the role of the $D$-state probability and short range nature of the nucleon-nucleon interaction. For that we consider two classes of interactions: (i)~The Yamaguchi and Yamaguchi~(YY)~\cite{YY54} separable potential with 4\% and 7\% $D$-state probability. Each has a different $D$-state probability and no short range repulsion. (ii)~The Unitary Pole Approximation~(UPA)~\cite{AR73,AR75} to the original Reid soft core potential (Reid68)~\cite{Re68} and the Nijmegen modified Reid potential (Reid93)~\cite{SKTS94}. The UPA potential by definition generates the same deuteron wave function as the original potential~\cite{AR75} that provided the optimum fit to the available data at the time the potentials were constructed and includes short range repulsion. In addition the models have different $D$-state probabilities for the deuteron.

For the Yamaguchi and Yamaguchi potentials~\cite{YY54} the form factor $g_\ell^\alpha(k)$ is given by
\begin{equation}
g_\ell(k) = \frac{k^\ell}{(k^2+\beta_\ell^2)^{(\ell+2)/2}}\ ,  \label{eq:19}
\end{equation}
where the parameters $\beta_\ell$ and $\lambda_{\ell\ell'}$ are detailed in Table~\ref{table:1}. Also included in this table are the binding energy $\epsilon_D$ and the quadrupole moment $Q_D$ for these two potentials. 

\begin{table}[h]
\caption{Parameters for the Yamaguchi-Yamaguchi potentials~\cite{YY54} with 4\% and 7\% 
$D$-state probability for the deuteron. Also included are the binding energy and quadrupole moments.}\label{table:1}
\vspace{0.3 cm}
\begin{tabular}{|l|c|c|c|c|c|c|c|} \hline
~$D$-state~ & $\beta_0$ & $\beta_2$ & $\lambda_{00}$ & $\lambda_{02}$ & $\lambda_{22}$ 
& $\epsilon_D$~(MeV) & $Q_D$\\ \hline
~~4\%  & ~1.3134~ &~ 1.5283~ &~ -0.6419 ~&~ 1.0849 ~&~ -1.8320 ~& 2.2234 &~ 0.2821 ~\\ \hline
~~7\% & 1.2410 & 1.9480 & -0.3776 & 1.6975 & -7.6301 & 2.2265  & 0.2826\\ \hline
\end{tabular}
\end{table}

\begin{table}[h]
\caption{The strength $\lambda_{\ell \ell'}$ for the UPA approximation to the Reid68~\cite{Re68}. and Reid93~\cite{SKTS94} 
potentials}\label{table:2}
\vspace{0.3 cm}
\begin{tabular}{|c||c|c|c|} \hline
 ~potential ~&$\lambda_{00}$ & $\lambda_{02}$ & $\lambda_{22}$ \\ \hline
~Reid68~ &~~-5.2896725E-02~~  &~~ -2.4385786E+00~~ &~~ 1.1850926E+00~~ \\ \hline
Reid93    & -4.7704789E-01 & -1.8111764E+00 & 2.5825467E-01 \\ \hline
\end{tabular}
\end{table}

In constructing the UPA to the Reid68~\cite{Re68} and Reid93~\cite{SKTS94} we have used the method of moments~\cite{AR75} to solve the Schr\"odinger equation for the deuteron wave function in coordinate space using the original potentials. This was achieved by taking the form factors such that the resultant deuteron wave functions for the Reid68 and Reid93 are linear combinations of the Yamaguchi-Yamaguchi type wave functions with different range parameters $\beta_i$, and therefore of the form
\begin{equation}
g_\ell(k) = \sum_{i=1}^{12}\ \frac{c_\ell^i\ k^\ell}{(k^2+\beta_i^2)^{(\ell+2)/2}}\ . \label{eq:20}
\end{equation} 
The strengths of the UPA potential ($\lambda_{\ell\ell'}$), adjusted to reproduce the matrix elements of the original Reid68 and Reid93 potentials, are given in Table~\ref{table:2}, while the parameters of the UPA form factors $\beta_i$ and $c^i_\ell$ for $\ell=0$ and $2$ are given in Table~\ref{table:3}. Here, we have chosen the range parameters $\beta_i$ to be multiples of the pion mass with the hope of reproducing some of the analytic structure of the one pion tail in the original Reid potentials.

\begin{table}[h]
\caption{The form factor parameters of the UPA approximation to the Reid68~\cite{Re68} and Reid93~\cite{SKTS94} potentials~.}\label{table:3}
\vspace{0.3 cm}
\begin{tabular}{|c|c||c|c||c|c|} \hline
       &           &\multicolumn{2}{|c||} {Reid68} & \multicolumn{2}{c|} {Reid93} \\ \cline{3-6}
$i$  &~ $\beta_i$ (fm$^{-1}$)~  & $c^i_0$ & $c^i_2$ & $c^i_0$ & $c^i_2$\\ \hline
 ~~1~~  & 0.7 &~ 7.21186419E-03 ~&~-2.24457073E-03~&~ 6.30646724E-03 ~&~ -3.08893140E-03 ~\\ \hline
 2  & 1.4 &   1.78826642E-01 & -3.31063031E-01  &  2.12846533E-01  &  -3.01564884E-01 \\ \hline
 3  & 2.1 & 1.31260692E+00  & -1.04745293E+00 &  6.05450638E+00 &  -1.78185516E+00 \\ \hline
 4  & 2.8 &  2.13430424E+00 & -1.43628043E+00 & -2.57777824E+01 &  7.87042755E-01  \\ \hline
 5  & 4.2 & 1.46578861E+02 &  -1.95695256E+01 & 3.20079733E+02  & -2.53483826E+01 \\ \hline
 6 & 5.6  &-8.10387728E+02 &  3.12782173E+00  & -1.49174373E+03 &  4.67387261E+01 \\ \hline
 7 & 7.0  & 1.12934549E+03 &  1.51126963E+02  &  2.32746050E+03 &  3.37908596E+01 \\ \hline
 8 & 9.8  &-5.87779728E+02 & -4.26701986E+02  & -2.57402658E+03 & -2.10353562E+02 \\ \hline
 9 &12.6 &-2.27638508E+02 &  5.92398037E+02  &  2.53223423E+03 &  3.41412020E+02 \\ \hline
 10&15.4& 5.33784864E+02 & -3.73533199E+02  & -1.31246553E+03 & -2.42126156E+02 \\ \hline
 11&21.0&-2.53746105E+02 &  9.68400708E+01  &  2.66329930E+02 &  7.60609941E+01 \\ \hline
 12&26.6& 6.63870056E+01 & -2.09513706E+01  & -4.84106437E+01 & -1.89979113E+01 \\ \hline
\end{tabular}
\end{table}

To establish the quality of the UPA deuteron wave function generated using the method of moments we present in Table~\ref{table:4} the deuteron properties for the original potential and the UPA for both Reid68 and Reid93. Also included are the effective range parameters to illustrate the domain of agreement in the scattering amplitude between the original and the UPA potential. It is clear from these results that the method of moments gives a very good representation of the original deuteron wave function and can reproduce the effective range parameters.

\begin{table}[h]
\caption{Comparison of the deuteron properties for the original potential and the UPA potential for both Reid68 and Reid93. Tabulated are the binding energy $\epsilon_D$, the asymptotic $S$-wave normalization $A_S$, the ratio  of the asymptotic $D$-wave to $S$-wave $\eta$, the quadrupole moment $Q_D$, and the $D$-state probability $P_D$. Also included are the scattering length $a_t$ and effective range $r_t$.}\label{table:4}
\vspace{0.3 cm}
\begin{tabular}{|c||c|c||c|c|}\hline
 & \multicolumn{2}{|c||} {Reid68} & \multicolumn{2}{c|} {Reid93} \\ \cline{2-5}
  & UPA & Original & UPA & ~Original~ \\ \hline
$\epsilon_D$ & 2.2246 & 2.2246 & 2.2246 & 2.2246 \\ \hline
$A_S$ & 0.87893 & 0.87758 & 0.8863 & 0.8853 \\ \hline
~~$\eta=A_D/A_S$ ~~&~ 0.026556 ~&~ 0.026223~ &~ 0.02565~ & 0.0251 \\ \hline
$Q_D$ & 0.2800 & 0.27964 & 0.2709 & 0.2703 \\ \hline
$P_D$ & 6.4691 & 6.4696 & 5.699 & 5.699 \\ \hline
$a_t$   & 5.408 & 5.390 & 5.445 & 5.422 \\ \hline
$r_t$   & 1.752 & 1.720 & 1.799 & 1.755 \\ \hline
\end{tabular}
\end{table}

\begin{figure}[h]
\caption{Comparison of the $^3$P$_1$ phase shifts for the Mongan potentials (Old) with Case I form factor and rank one (R=1) and rank two (R=2) with the new fit (New) and the experimental ($Exp.$) Nijmegen~\cite{Nij93} $np$ data. }\label{fig1}
\vspace{0.3 cm}
\centering\includegraphics[scale=0.6]{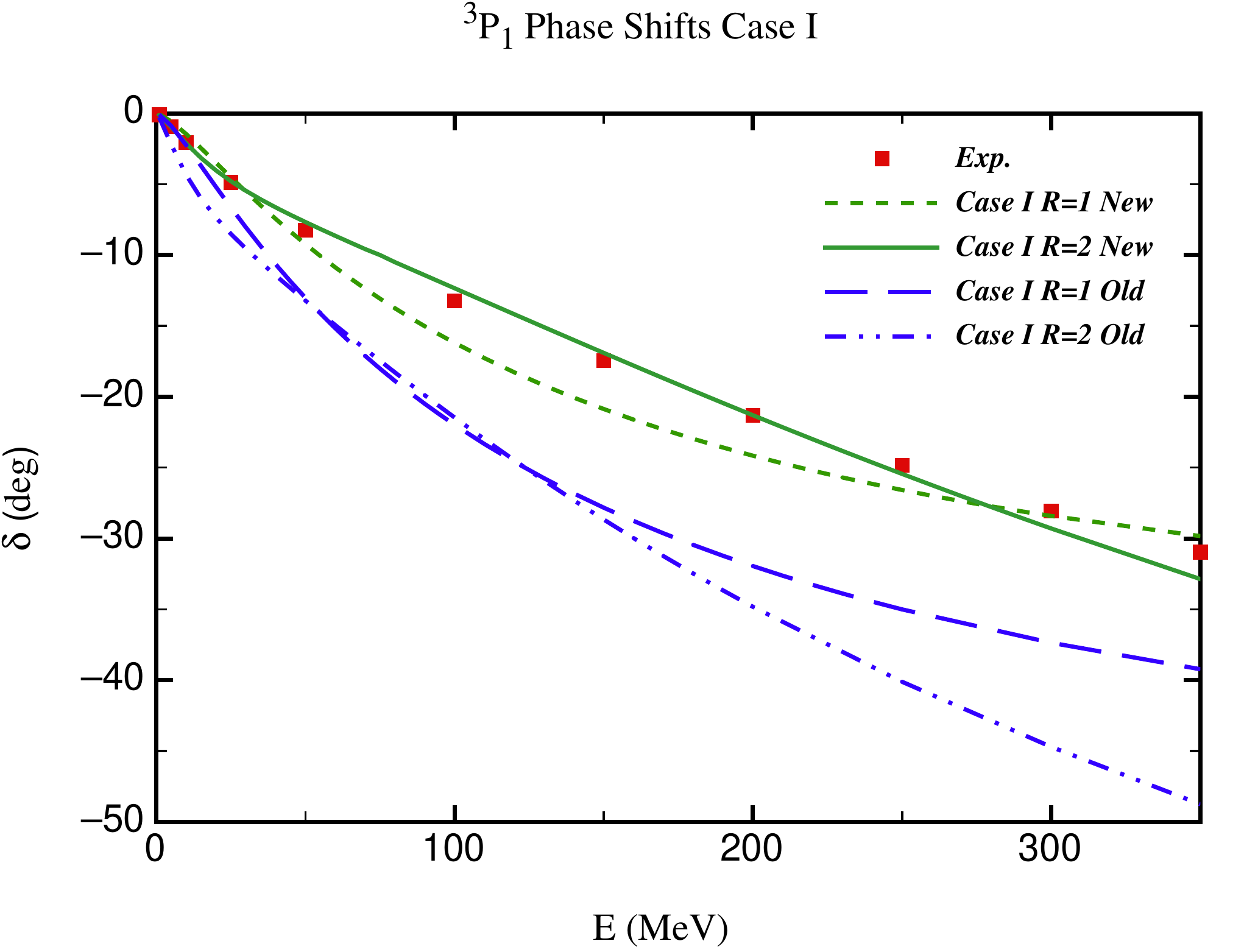}
\end{figure}

Finally, to examine the importance of multiple scattering in determining the deuteron EDM, we need to introduce a $^3$P$_1$ interaction to calculate $d_{MS}$. Here we need to know how important is the fit to the data and the role of the off-shell amplitude in determining the magnitude of $d_{MS}$. To simplify the evaluation of $d_{MS}$, we have chosen to use separable potentials with different form factors. The Mongan~\cite{Mo69} potentials used by Avishai~\cite{Av85} come with different form factors, and therefore different off-shell properties. They are either rank one or rank two to optimize the fit to the data; \textit{i.e.}, the potentials are of the form
\begin{equation}
v_{^3P_1}(k,k') = \sum_{i=1}^n\ g_i(k)\,\lambda_i\,g_i(k')\ ,\label{eq:21}
\end{equation}
where $n=1$ for rank-one potentials and $n=2$ for rank-two potentials. For the form factors $g_i(k)$ we will use the four different forms chosen by Mongan (see Table~\ref{table5}). Considering the fact that Mongan adjusted the parameters of his potentials to fit the Livermore data of the 1960's, we need first compare the phase shifts predicted by the Mongan potentials and those that we constructed to fit the latest Nijmegen~\cite{Nij93} $np$ data. In Fig.~\ref{fig1} we compare the $^3$P$_1$ phase shifts for rank-one and rank-two Case I form factors for Mongan's potentials with those refitted to the Nijmegen data. Also included are the Nijmegen~\cite{Nij93} $np$ phase shifts. It is clear from the the results in Fig.~\ref{fig1} that the original Mongan potentials give a poor fit to the current data, while the new fits reproduce the data to a much better degree. Since the $^3$P$_1$ amplitude required for the determination of $d_{MS}$ is evaluated at the deuteron binding energy, \textit{i.e.}, \textit{below} the elastic threshold, it is essential that we fit well the low energy phase shifts. Because these are small, we have chosen the criteria for a good fit $\chi^2$ defined as
\begin{equation}
\chi^2 = \sum_{i=1}^{n}\ \frac{|\delta_i^{\rm th}-\delta_i^{\rm exp}|^2}{|\delta_i^{\rm exp}|^2}\ ,\label{eq:22}
\end{equation}
where $n=11$ is the number of data points below 300~MeV. In Table~\ref{table5} we present new fits to the Nijmegen $np$ data for the different form factors used by Mongan~\footnote{The Case III form factor was motivated by the observation that the on-shell Born amplitude for a rank-one separable potential is identical to the on-shell Born amplitude resulting from meson exchange potential with a meson mass $\beta_1$.}. Included are rank-one and rank-two potentials and the $\chi^2$ for each potential. It is clear from the $\chi^2$ that the rank two potentials give a better fit. This is especially true for the Case I form factor. In the following discussion of the deuteron EDM we will consider these different $^3$P$_1$ potentials to establish the importance of fitting the data and the role of the off-shell behavior of the amplitude.

\begin{table}[h]
\caption{The parameters of the `New' rank-one and rank-two potentials with the different Mongan form factors. The parameters are adjusted by minimizing the $\chi^2$ defined in Eq.~(\ref{eq:22}) taking the experimental phases from the latest Nijmegen~\cite{Nij93} $np$ phase shift analysis. The form factor for Case III is written in terms of $Q_1(\xi)$ the Legendre function of the second kind. }
\label{table5}
\vspace{0.3 cm}
\begin{tabular}{|l|c|c|c|c|c|c|c|}\hline
Potential ~& form factor $g_i(k)$ &~Rank~ & $\beta_1$  & $\lambda_1$ & $\beta_2$ & $\lambda_2$ & $\chi^2$  \\ \hline
 Case I     & $k/(k^2+\beta_i^2)$ & 1 &~1.725~ & 0.95 & - & - & 0.62  \\ \cline{3-8}
               &                               & 2 & 0.90 & ~ 0.059~ &~ 3.58~ & ~-2.0~  & ~0.02~ \\ \hline
 Case II   &$k/(k^2+\beta_i^2)^{3/2}$ & 1 & 2.38 & 9.35 &-  & - & 0.81   \\ \hline
 Case III   &~$\left[\frac{1}{k^2\pi} Q_1(1+\frac{\beta_i^2}{2k^2})\right]^{1/2}~$ & 1 & 1.68 & 60.0 &- &- & 0.19  \\ \cline{3-8}
  & &  2 & 1.20 & 120.0 & 4.4 & -2.3 & 0.12 \\ \hline
 Case IV  &$k/(k^2+\beta_i^2)^{2}$ & 1& 2.715 & 147.0 & -& -& 0.78  \\ \hline
\end{tabular}
\end{table}

\subsection{The deuteron EDM}

We now turn to the study of the sensitivity of the deuteron EDM to the nuclear structure effects as defined by the strong nucleon-nucleon interactions detailed above. We first consider the sensitivity of the two-body deuteron EDM $d_D^{(2)}$ to the $D$-state 
probability ($P_D$). In Table~\ref{table6} we summarize the contributions to the deuteron 
EDM for the four different deuteron wave functions being considered. For the $^3$P$_1$
interaction we use a rank-two Mongan Case I potential (fitted to the latest Nijmegen
phase shifts~\cite{Nij93}).  Also included are the results of Khriplovich and 
Korkin~\cite{KK00}.   We observe that in the plane wave approximation ($d_{PW}$) there 
is little variation with $P_D$, and the short range repulsion incorporated in the two Reid
potential wave functions provides no more than a 10\% reduction in $d_{PW}$.  Moreover, 
the results are effectively consistent with the zero range (chiral limit) approximation of 
Khriplovich and Korkin.  In particular, the plane wave results for the two YY models
suggest that the dependence upon the deuteron $D$-state probability is such that an
S-state deuteron result would approach that of Ref.~\cite{KK00}.  In contrast, the multiple 
scattering contribution ($d_{MS}$), which is of the opposite sign to the plane wave term, varies 
considerably depending upon the short range character of the deuteron wave function.  
In particular, the two Reid potentials with different $P_D$ values yield quite similar values of 
$d_{MS}$, but these are only half those generated by the YY potentials.  The difference
between the YY and Reid potential models can be understood in light of our knowledge
that there is no explicit short range repulsion in the YY potentials. We will return to this difference when we address the role of the off-shell behavior of the $^3$P$_1$ amplitude in determining the magnitude of the multiple scattering contribution $d_{MS}$. From these results we may conclude that the 
strong repulsion at short distance in realistic nucleon-nucleon potentials reduces the 
effects of multiple scattering in the matrix element to such an extent that the multiple 
scattering contribution $d_{MS}$ is only about 20\% of the plane wave contribution 
$d_{PW}$.  Furthermore, as noted above, the final results are not particularly sensitive 
to $P_D$.
 
\begin{table}[h]
\caption{The variation of the two-body EDM with $D$-state probability of the deuteron. For the $^3$P$_1$ interaction we use the `New' fit Case I rank-two potential.
Also included are the results of Khriplovich and Korkin~\cite{KK00}.}
\label{table6}
\vspace{0.3 cm}
\begin{tabular}{|l|c|c|c|c|}\hline
$^3$S$_1$-$^3$D$_1$  & $P_d$ & $d_{PW} (A\,e\,fm)$ & $d_{MS} (A\,e\,fm)$ & $d_D^{(2)} (A\,e\,fm)$ \\ \hline
YY 4\%   & 4\% & -1.035 & 0.4115 & -0.6234 \\ \hline
Reid93    & 5.7\% & -0.9715 & 0.2009 & -0.7706 \\ \hline
Reid68   & 6.5\% & -0.9620 & 0.1718 & -0.7902 \\ \hline
YY 7\% & 7\% & -1.083 & 0.4271 & -0.6564 \\ \hline
Khriplovich \textit{et al.} & & -0.92 & & \\ \hline
\end{tabular}
\end{table}

To establish the importance of the multiple scattering contribution ($d_{MS}$) to the total two-body deuteron EDM, we turn to the dependence of $d_D^{(2)}$ on the choice of the $^3$P$_1$ interaction. But first we need to examine the sensitivity of the multiple scattering contribution to the $^3$P$_1$ phase shifts. This can be achieved by comparing the results for the EDM using the `Old'  Mongan fit to the 1960's Livermore phase shift analysis  and the `New' fit with the same separable potential form factors to the latest Nijmegen~\cite{Nij93} $np$ data. We have in Table~\ref{table7} the EDM results for the rank-one separable potentials with Case I and III form factors. For the deuteron wave function we have used either the UPA to the Reid68 or the YY 4\% potentials. It is clear from these results that the multiple scattering contribution ($d_{MS}$) is reduced as a result of the fit to the more recent phase shift analysis (compare rows four and five or rows six and seven in Table~\ref{table7}). This reduction in $d_{MS}$ is consistent with the observation that the `New'  $^3$P$_1$ potentials provide less repulsion (i.e. smaller phase shifts, see Fig.~\ref{fig1}) and, therefore, substantially smaller multiple scattering contributions than the old fits due to Mongan. This observation is encouraging for extending the above analysis based on $d_{PW}$ to the three-nucleon EDM, as the new $np$ data suggest a reduced contribution from the multiple scattering term. 

We now return to the role of the short range repulsion in the deuteron wave function on the magnitude of the multiple scattering term $d_{MS}$ as illustrated in Table~\ref{table6}. In comparing the results for the Reid68 and 4\%~YY deuterons (column three and five in Table~\ref{table7}) for the Case~I and Case~III $^3$P$_1$ potentials, we find that the multiple scattering term is suppressed  for both $^3$P$_1$ potentials. This suggests that the effect tabulated in Table~\ref{table6} might be valid in general. which implies that the inclusion of multiple scattering will require a more realistic treatment of the deuteron wave function than is the case for the zero range approximation employed by Khriplovich and Korkin~\cite{KK00}. In fact for some combination of deuteron wave function and  $^3$P$_1$ interaction ( 4\% YY and Case III Old) the multiple scattering contribution ($d_{MS}$) is about the same size as the plane-wave approximation ($d_{PW}$) and as a result the deuteron EDM $d_D^{(2)}$ is suppressed by an order of magnitude compared to the combination Reid68 and Case I New.

\begin{table}[h]
\caption{Variation in the deuteron EDM with changes in the $np$ phase shifts for two rank-one 
separable potentials having different form factors as defined by Mongan~\cite{Mo69}. Here 
`New' refers to the fit to the latest Nijmegen~\cite{Nij93} $np$ phase shifts while `Old' refers to the 
original Mongan fit. }
\label{table7}
\vspace{0.3 cm}
\begin{tabular}{|c|c||c|c||c|c|}\hline
\multicolumn{2}{|c||}{$^3$S$_1$-$^3$D$_1$} &\multicolumn{2}{c||}{Reid68} &\multicolumn{2}{c|}{YY 4\%} \\ \hline
\multicolumn{2}{|c||}{}&\multicolumn{2}{c||}{~$d_{PW}=-0.96$~~} &\multicolumn{2}{c|}{~$d_{PW}=-1.04$~~} \\ \hline
Case  & $\chi^2$ &~ $d_{MS}$~ & $d_D^{(2)}$ & ~$d_{MS}$~ & $d_D^{(2)}$ \\ \hline
I (New) &~ 0.62~ & 0.21 &-0.75 &0.57 &-0.47  \\ \hline
I (Old) & 1.90 & 0.31 & -0.66 & 0.78 & -0.26 \\ \hline
III (New) & 0.19 &0.25 & -0.71 & 0.77 & -0.27 \\ \hline
III (Old) &6.67 & 0.42 & -0.54 & 1.16 & 0.12 \\ \hline 
\end{tabular}
\end{table}

\begin{table}[h]
\caption{The dependence of $d_{MS}$ on the $^3$P$_1$  separable potential form factor as 
defined by Mongan~\cite{Mo69} and that are fit to the latest $np$ phase shifts. The Reid93 
or the 4\% YY deuteron wave function is used in all cases as indicated.}
\label{table8}
\vspace{0.3 cm}
\begin{tabular}{|c|c|c||c|c||c|c|}\hline
 & & & \multicolumn{2}{c||}{ Reid93 } & \multicolumn{2}{c|}{ YY 4\%} \\ \cline{4-7}
  & & & \multicolumn{2}{c||}{$d_{PW}=-0.9715$} & \multicolumn{2}{c|}{$d_{PW}= -1.035$} \\ \cline{4-7}
Case &~Rank~&~ $\chi^2$~ &$~d_{MS} (A\,e\,fm)$ ~&~ $d_D^{(2)} (A\,e\,fm)$ ~&
$~d_{MS} (A\,e\,fm)$ ~&~ $d_D^{(2)} (A\,e\,fm)$ ~\\ \hline
 I   & 1 &~ 0.62~  &0.2583 & -0.7132  & 0.5665 & -0.4684 \\ 
 I  & 2 & 0.02  &0.2009 & -0.7706 &0.4115 & -0.6234 \\ \hline
 II & 1 & 0.81  &0.2229 & -0.7486 &0.3807 & -0.6542 \\ \hline
 III & 1 & 0.19  &0.3075 & -0.6640 &0.7654 &-0.2696 \\
 III & 2 & 0.12  &0.3805 & -0.5910 &1.108 & 0.0734 \\ \hline
 IV & 1 & 0.78  &0.2153 & -0.7562 & 0.3277 &-0.7072 \\ \hline
\end{tabular}
\end{table}

We now turn to the role of the off-shell behavior of the $^3$P$_1$ amplitude in the deuteron EDM. Here again we make use of the different separable potentials with the different form factors used by Mongan after readjusting the parameters of the potential  to fit the latest Nijmegen~\cite{Nij93} $np$ phase shifts. The parameters of these `New' potentials are given in Table~\ref{table5}. In Table~\ref{table8} we report the multiple scattering contribution $d_{MS}$ and the two-body EDM $d_D^{(2)}$  for these separable potentials.  In each case we have made use of either the Reid93  or 4\% YY deuteron wave functions in the calculations. Here we observe that for the Reid93 deuteron there is a smaller variation in $d_D^{(2)}$  than is the case for the 4\% YY deuteron. This is due to the fact the the multiple scattering contributions, $d_{MS}$, for the 4\% YY deuteron have a substantially larger variation for the different fits to the $np$ data. This is consistent with the results in Table~\ref{table7} and is due to the absence of short range repulsion in the YY potentials. Here we can raise a number of questions regarding the role of the $^3$P$_1$ amplitude in determining the magnitude of the multiple scattering contribution $d_{MS}$. These are:
\begin{itemize}
\item Why is $d_{MS}$ almost a factor of two smaller for the Reid93 when compared to that for the 4\%~YY potential?
\item Why, for the Reid93 deuteron, is $d_{MS}$ about the same for all form factors with the possible exception of Case III which gives the largest contribution? 
\item Why is it that for the 4\%~YY deuteron $d_{MS}$ has a much larger variation than is the case for Reid93?
\end{itemize}
 To address these questions and to try to correlate the results in Table~\ref{table8} with the off-shell behavior of the $^3$P$_1$ amplitude, we need to examine the analytic continuation of the $P$-wave scattering wave function to the deuteron pole. This is defined in momentum space in terms of the half off-shell $t$-matrix as
\begin{eqnarray}
\psi_\alpha(k) &=& G_0(-\epsilon_D,k)\ t_\alpha(k,i\kappa;-\epsilon_D) \nonumber \\
&=& G_0(-\epsilon_D,k)\ \mathbf{g}_\alpha(k)\,\bm{\tau}_\alpha(-\epsilon_D)\,\mathbf{g}_\alpha^\dag(i\kappa)\nonumber \\
&\equiv& \Psi_\alpha(k)\,\bm{\tau}_\alpha(-\epsilon_D)\,\mathbf{g}_\alpha^\dag(i\kappa)\ , \label{eq:23}
\end{eqnarray}
where $\alpha$ labels the $^3$P$_1$ channel, $\epsilon_D =\frac{\kappa^2}{2\mu}$ is the binding energy of the deuteron, and $\mu$ is the $np$ reduced mass.   Here the free Green's function at the deuteron energy is given by $G_0(-\epsilon_D,k) = -(2\mu)(\kappa^2+k^2)^{-1}$, while the amplitude $t_\alpha(k,i\kappa;-\epsilon_D)$ is the half off-shell $^3$P$_1$ $t$-matrix evaluated at the deuteron pole. In the second line of Eq.~(\ref{eq:23}) we have written the off-shell $t$-matrix in its separable form with
\begin{equation}
\bm{\tau}_\alpha(-\epsilon_D) = \left[\,\bm{\lambda}_\alpha +2\mu \int\limits_0^\infty dk\,k^2\ \frac{\mathbf{g}_\alpha^\dag(k)\,\mathbf{g}_\alpha(k)}{\kappa^2 + k^2}\,\right]^{-1}\ .\label{eq:24}
\end{equation}
For rank-one potentials $\bm{\tau}_\alpha(-\epsilon_D)$ is positive definite since the potential is repulsive (\textit{i.e.}, $\lambda_\alpha>0$). As a result, the scattering wave function can be written as
\begin{equation}
\psi_\alpha(k) = \chi_\alpha(k)\ \sqrt{\tau_\alpha(-\epsilon_D)}\ g_\alpha(i\kappa)\ .\label{eq:25}
\end{equation}
This definition of the function $\chi_\alpha(k)$ is motivated in the following discussion of the matrix elements of the dipole operator $O_d$ and the $PT$ violating one pion exchange potential $V$ that go into the evaluation of $d_{MS}$.

\begin{figure}[h]
\caption{Comparison of the $^3$P$_1$ `scattered function' $\chi_\alpha(k)$ defined in Eq.~(\ref{eq:25}) for the rank-one separable potentials that fit the latest Nijmegen~\cite{Nij93} $np$ phase shifts.}\label{fig2}
\vspace{0.3 cm}
\centering\includegraphics[scale=0.6]{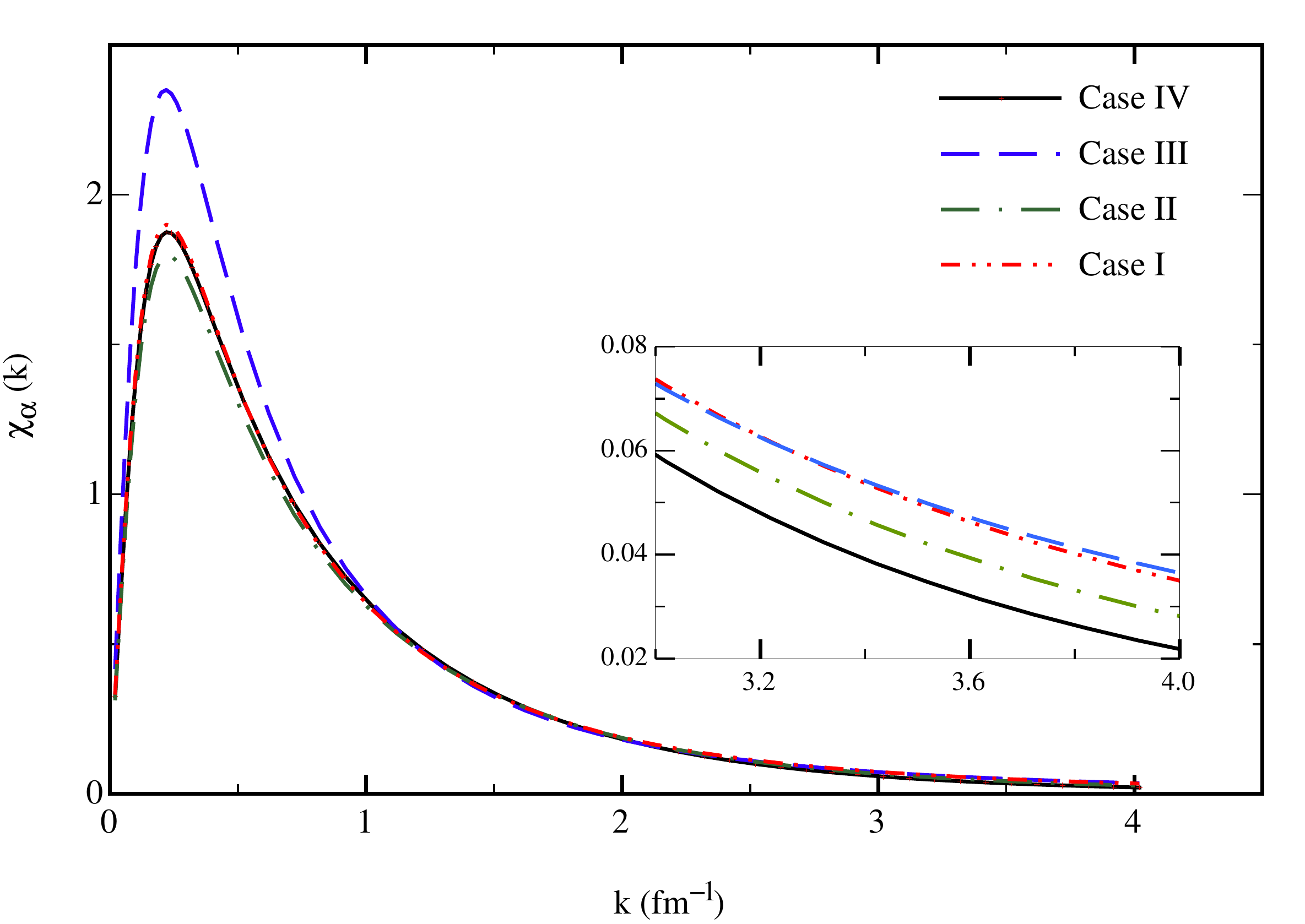}
\end{figure}

In Fig.~\ref{fig2} we plot the function $\chi_\alpha(k)$ for all the rank-one potentials used in Table~\ref{table8}. A careful inspection of this figure reveals that: (i)~ The `scattered function' $\chi_\alpha(k)$ for the Case III form factor is substantially larger for $k<1.0$~fm$^{-1}$ than that of the other three form factors. (ii)~For $k>3$~fm$^{-1}$ the Case III `scattered function' has the longest range followed by Case I and then Case II and finally Case IV. This is clear from the choice of form factors as given in Table~\ref{table5}.

To establish how this momentum dependence of the $^3$P$_1$ `scattered function' effects the multiple scattering contribution $d_{MS}$ to the deuteron EDM, we recall from Eq.~(\ref{eq:14}) that $d_{MS}$ can be written as
\begin{equation}
d_{MS} = -2\left[\,\mathbf{O}_{sp}+\mathbf{O}_{dp}\,\right]\,\bm{\tau}(-\epsilon_D)\,
\left[\,\mathbf{V}_{ps}+\mathbf{V}_{pd}\,\right]\ ,\label{eq:26}
\end{equation}
where
\begin{equation}
\mathbf{O}_{D \alpha} = \bra\Psi_D|\,O_d\,|\Psi_\alpha\ket\quad\mbox{and}\quad
\mathbf{V}_{\alpha D} = \bra\Psi_\alpha|\,V\,|\psi_D\ket\ ,\label{eq:27}
\end{equation}
with $D=\ ^3$S$_1$ or $^3$D$_1$ and $\alpha=^3$P$_1$. For rank-one separable potentials we can absorb a factor of $\sqrt{\tau(-\epsilon_D)}$ into the matrix elements, i.e. ${\cal O}_{\alpha p}\equiv O_{\alpha\beta}\sqrt{\tau(-\epsilon_D)}$ and ${\cal V}_{p \alpha} \equiv \sqrt{\tau(-\epsilon_D)}\,V_{p \alpha}$ and therefore for rank-one potentials we have
\begin{equation}
d_{MS} = -2\left[\,{\cal O}_{sp} + {\cal O}_{dp}\,\right]\ \left[\,{\cal V}_{ps} + {\cal V}_{pd}\,\right]\ .\label{eq:28}
\end{equation}
The values of ${\cal O}_{\alpha p}$ and ${\cal V}_{p\alpha}$ for the four different form factors and with a deuteron wave function given by either the 4\% YY or the Reid93 are presented in Table~\ref{table9}. It is clear from these results that the matrix elements of the dipole operator $O_d$, which is long range in coordinate space, are to a good approximation independent of the deuteron wave function and to within 20\% independent of the $^3$P$_1$ potential. On the other hand the matrix elements of the $PT$ violating one pion exchange potential, which probes the short range behavior of both the $^3$P$_1$ and the deuteron wave function, are clearly model dependent. In particular, for the Reid93 deuteron with short range repulsion, the variation in ${\cal V}_{p\alpha}$ is small with the Case III form factor giving the largest contribution and Case IV yielding the smallest contribution followed by Case II and Case I. This is consistent with the observation made in the Fig.~\ref{fig2} insert regarding the asymptotic behavior of the function $\chi_\alpha(k)$.  This is also consistent with the observation in Table~\ref{table8} for rank-one potentials. On the other hand, for the 4\% YY deuteron, with no short range repulsion, the matrix elements are almost a factor of two larger  with the Case III form factor giving the largest contribution and Case IV the smallest. From the results in Table~\ref{table9} we may conclude that it is the matrix element of the $PT$ violating one pion exchange potential that probes the short range behavior of the $^3$P$_1$ and deuteron wave functions and, as a result, determines the magnitude of $d_{MS}$. To that extent it is essential that one generate those two wave functions in a consistent frame work. On the other hand, when the deuteron includes the short range behavior dictated by modern nucleon-nucleon interactions, the contribution of the multiple scattering term $d_{MS}$ is suppressed ($\approx 20$\%) in comparison to the plane wave contribution $d_{PW}$. This suggests that one may be able to evaluate the EDM for the three-nucleon system in the plane wave approximation in such a model with an error of the order of 20\%.

\begin{table}[h]
\caption{The matrix elements of the dipole operator $O_d$ and the $PT$ violating one pion exchange potential $V$ for the four different form factors and two different deuteron wave functions.}
\label{table9}
\vspace{0.3 cm}
\begin{tabular}{|c|c||c|c||c|c|}\hline
~deuteron ~&~ Case ~& ${\cal O}_{sp}$ & ${\cal O}_{dp}$ & ${\cal V}_{ps}$ & ${\cal V}_{pd}$ \\ \hline
4\%~YY & I       &~ -0.4197 ~&~ -0.05599 ~&~ 0.5533 ~&~ 0.04211 ~ \\ \cline{2-6}
              & II     & -0.4039  & -0.05422 & 0.3794 & 0.03618 \\ \cline{2-6}
              & III    & -0.4819 & -0.06300& 0.6578 & 0.04444 \\ \cline{2-6}
              & IV   & -0.4185 & -0.05622 & 0.3124 & 0.03276 \\ \hline
  Reid93  & I     & -0.4221 & -0.06069 & 0.2169 & 0.09031 \\ \cline{2-6}
              & II    & -0.4068 & -0.05906 & 0.1928 & 0.04641 \\ \cline{2-6}
              & III   & -0.4852 & -0.06712 & 0.2269 & 0.05154 \\ \cline{2-6}
              & IV   & -0.4224 & -0.06105 & 0.1793 & 0.04338 \\ \hline 
\end{tabular}
\end{table}

Finally, the results in Table~\ref{table9} for ${\cal O}_{\alpha p}$ and ${\cal V}_{p\alpha}$ indicate that the contribution from the $D$-wave component of the deuteron wave function are an order of magnitude smaller than the $S$-wave component. This may suggest that one could neglect the $D$-wave component in the calculating $d_{MS}$ and a simplification of the calculation of the multiple scattering term in heavier nuclei. This observation is consistent with the results in Table~\ref{table6} where the changes in the multiple scattering contribution has a variation of about 10\% with $D$-state probability.

\section{Conclusions}

From our analysis we offer the following conclusions: (i)~In the absence of multiple scattering ($d_{MS}=0$) 
the variation in $d^{(2)}_D$ due to differences in the deuteron wave functions is less than 5\%, and the value 
of $d_{PW}$ is consistent with the zero range (chiral limit) results of Khriplovich and Korkin~\cite{KK00}. 
(ii)~The contribution from multiple scattering  $d_{MS}$ is sensitive to the short range behavior of the 
deuteron wave function, and the $d_{MS}$ contribution is about 20\% for realistic parametrizations of the 
deuteron such as those represented by the Reid93 potential model. This suggests that we can extend the 
analysis to heavier nuclei in the plane wave approximation with an estimated error of $\approx 20$\%. 
(iii)  As suggested by Liu and Timmermans, one pion exchange dominates the deuteron EDM calculation.
(iv)~The contribution from the $^3$P$_1$ interaction via $d_{MS}$ depends on the phase shifts in this 
channel as well as the off-shell behavior of the amplitude.  (v)~A comparison of our Reid93 results with those of Liu and Timmermans~\cite{LT04} 
indicates that one can use a separable potential approximation in heavier nuclei, \textit{e.g.}, $^3$He and 
$^3$H, with minimal loss in accuracy.  Moreover, until deuteron EDM experiments attain an uncertainty of 
less than 10\%, simple separable potential model calculations should provide an adequate description.

\section{Acknowledgement}
The work of BFG was performed under the auspices of the National Nuclear Security Administration of the 
U.S. Department of Energy at Los Alamos National Laboratory under Contract No DE-AC52-06NA25396.

\newpage



\end{document}